\documentclass [12pt,preprint]{aastex}

\begin{document}

\title{Evidence for Accretion: \\
High-Resolution X-ray Spectroscopy of the Classical T
Tauri Star TW Hydrae}

\author{Joel H. Kastner \\
\small
Chester F. Carlson Center for Imaging Science, Rochester Institute of 
Technology, 54 Lomb Memorial Dr., Rochester, NY 14623; jhk@cis.rit.edu \\
\normalsize
David P. Huenemoerder, Norbert S. Schulz, Claude R. Canizares \\
\small
MIT Center for Space Research, 70 Vassar St., Cambridge, MA, 02139,
U.S.A. \\
\normalsize
\and
David A. Weintraub \\
\small
Dept.\ of Physics and Astronomy, Vanderbilt University,
Nashville, TN, 37235, U.S.A.}

\begin{abstract}
  We present high resolution X-ray spectra of the X-ray bright
  classical T Tauri star, TW~Hydrae, covering the
  wavelength range of 1.5-25 \AA.   
  The differential emission measure derived from fluxes of
  temperature-sensitive emission lines shows a plasma with a
  sharply peaked temperature distribution, peaking at $\log T = 6.5$.  
  Abundance anomalies are apparent, with iron very deficient
  relative to oxygen, while neon is enhanced relative to oxygen.
  Density-sensitive line ratios of Ne~{\sc ix} and O~{\sc vii} indicate
  densities near $\log n_e = 13$. A flare with rapid ($\sim1$
  ks) rise time was detected during our 48 ksec observation; however,
  based on analysis of the emission-line spectrum during
  quiescent and flaring states, the derived plasma parameters
  do not appear strongly time-dependent. 
  The inferred plasma temperature distribution and densities are
  consistent with a model in which the bulk of the X-ray
  emission from TW Hya is generated via mass accretion from its
  circumstellar disk. Assuming accretion powers the X-ray
  emission, our results for $\log n_e$ 
  suggest an accretion rate of $\sim10^{-8} M_\odot$ yr$^{-1}$. 
\end{abstract}


\section{Introduction}

Over the past two decades --- beginning with {\it Einstein}
and continuing through the 
most recent observations by the {\it Chandra X-ray Observatory} and the X-ray
Multi-Mirror Mission (XMM) --- X-ray observatories have
produced an increasingly 
detailed and comprehensive census of X-ray sources in star formation
regions. These observations have firmly established
the presence of strong X-ray emission as one of the defining 
characteristics of stellar youth (Feigelson \& Montmerle
1999). Moreover, high-energy emission from young stars is central to many,
seemingly disparate aspects of star and planet formation, from the
chemical evolution of dark clouds to formation of chondrules found in
meteoritic inclusions.

Despite considerable recent
progress, astronomers are still grappling
with a fundamental understanding of the physical mechanisms
that lead to X-ray emission from
regions of young stars. For deeply convective T Tauri stars (TTS),
coronal activity due to intense surface 
magnetic fields has long been proposed as the 
source of the X-rays 
(Feigelson \& Montmerle 1999 and references therein). A coronal origin
appears plausible for weak-lined T Tauri stars
(wTTS), if the hypothesis is correct that these stars have
have either lost or become decoupled from their 
circumstellar accretion disks and generally are 
rotating more rapidly than the disk-enshrouded classical T Tauri Stars
(cTTS). Coronal activity may not be the only possible source
of X-rays, however. There is reason on theoretical grounds
to expect X-ray emission from star-disk interactions (e.g.,
Shu et al.\ 1997). Such interactions likely persist at least through
the cTTS evolutionary stage. 

Existing X-ray data are inconclusive as to the  
relation of TTS coronae to the coronae of main sequence
and evolved stars,
and to the potential distinctions between
X-ray emission from coronae vs.\ from star-disk interactions (e.g., Feigelson
2001). Moderate resolution ($R\sim50$)
CCD X-ray spectroscopy by {\it ASCA} provides tantalizing evidence
for abundance anomalies in the X-ray spectra of TTS (e.g.,
Skinner et al.\ 1997;  Skinner \& Walter 1998) that are 
similar to those seen in
{\it ASCA} CCD spectroscopy of active, evolved stars
(e.g., White 1996). These same studies also make clear, however, 
the limitations of stellar X-ray spectroscopy with CCDs. The
High Energy Transmission Gratings Spectrometer (HETGS) on
{\it Chandra} now provides a much
more powerful tool for the study of X-ray emission from
stars (Canizares et al.\ 2000). 
With a resolution of $\sim 1000$, Chandra/HETGS can easily
resolve individual spectral
lines and line complexes, providing
an arsenal of greatly improved plasma diagnostics (e.g.,
Drake et al.\ 2001). 

The TW Hydrae Association (TWA; Kastner et al.\ 1997, Webb
et al.\ 1999, Zuckerman et al.\ 2001) holds great potential for
spectral studies of 
X-ray emission from TTS. Due to the proximity of the
association ($D\sim50$ pc) and the lack of intervening, 
ambient cloud material, the X-ray flux of a
{\it typical} TWA star ($F_x \sim3\times10^{-12}$ erg cm$^{-2}$ s$^{-1}$)
rivals or exceeds that of the most
X-ray luminous TTS in dark clouds (Kastner et al.\
1997). As the X-ray spectra of TWA stars can be obtained
free of contamination from absorbing 
dark cloud material, any line-of-sight absorption 
can be attributed almost entirely to circumstellar gas. 

To date, the only TWA star that has been studied
in detail at X-ray wavelengths is TW Hya itself (Kastner et
al.\ 1999). At a distance of only 57 pc from Earth (Wichmann
et al.\ 1998), TW Hya is the nearest known classical TTS, and
is one the best studied examples of a young star-disk
system; its dust-rich, gaseous accretion disk evidently is 
viewed nearly face-on (Kastner et al.\ 1997; Krist et al.\
2000; Wilner et al.\ 2000; Weintraub, Kastner \& Bary 2000;
Trilling et al.\ 2001).
Based on ASCA and ROSAT spectral data, the X-ray emission
from TW Hya was modeled by Kastner et al.\ as arising in a two-component
plasma with $T_{x,1} \simeq 2$ MK and $T_{x,2} \simeq 10$
MK. The emission evidently is attenuated by a
variable absorbing column, with $N_H$ ranging from $\sim$5 $\times$ 
10$^{20}$ cm$^{-2}$ to $\sim$3 $\times$ 
10$^{21}$ cm$^{-2}$. From this modeling, Kastner et
al. (1999) concluded that the strong emission from TW Hya at
$\sim1$ keV (Fig.\ 1, top) probably was dominated by lines of highly
ionized Fe.

Here, we present high-resolution
Chandra/HETGS spectroscopy of TW Hya. These data dramatically
confirm that the X-ray emission from TW Hya is dominated
by emission lines --- although Fe emission is suprisingly weak and Ne
emission surpringly strong. Furthermore, Chandra/HETGS
spectra call into question a coronal origin for the X-ray
emission from TW Hya. 

\section{Observations and Results}

TW Hya was observed with Chandra/HETGS for 48 ks in June, 2000
(observation identifier 5) in the default configuration (timed
exposure, ACIS-S detector array), under nominal operating conditions.
Data were re-processed with CIAO software to apply updated
calibration, and events were cleaned of the detector artifacts on CCD
8 (``streaks'').  Spectral responses were made with CIAO
(Chandra Interactive Analysis of Observations) software.
Lines were measured with ISIS\footnote{ISIS is available from {\tt
    http://space.mit.edu/CXC/ISIS}} (Houck \& DeNicola 2000), and emission
measure and abundances were modeled with custom IDL software.  More
detailed descriptions of the processing can be found in
Huenemoerder, Canizares, \& Schulz (2001), in which the same
techniques were applied to HETGS spectra of II~Pegasi.

The resulting Chandra/HETGS spectrum yielded 3700 first
order medium energy grating (MEG) counts, and 
1080 first order high energy grating (HEG) counts. 
We present the TW~Hya spectrum in
Fig.\ 1 (bottom). The Chandra/HETGS data 
confirm that the X-ray spectrum of
TW Hya longward of $\sim10$ \AA\ ($E<1.2$ eV) is dominated
by emission lines. 
The measured flux (0.45-6.0 keV) is
$3.7\times10^{-12}\,\mathrm{ergs\,cm^{-2}\,s^{-1}}$ 
($2.5\times10^{-3}\,\mathrm{photons\,cm^{-2}\,s^{-1}}$), corresponding to
$L_x=1.4\times10^{30}\,\mathrm{ergs\,s^{-1}}$ at the
distance to TW Hya ($D=57$ pc). 
Because the HETGS response matrix is
nearly diagonal, this flux result is obtained independent of
a model spectrum. 

The Chandra/HETGS spectrum demonstrates
that the $\sim1$ keV
region of the spectrum is dominated by lines of {\it highly
ionized Ne} --- not Fe (Fig.\ 2). Furthermore, 
lines of ionized Fe are conspicuous for
their absence across much of the HETGS spectral
regime; instead, besides Ne, the most prominent lines are
due to O, Mg, and Si (Table 1). 

About 30 ks into the observation, a flare occurred during
which the flux doubled in about 1 ks. The flux immediately began
a rather linear decay, followed by a possible minor flare
(Fig.\ 3). We
inspected lightcurves in narrower bands and in lines for
indications of similar variability. Harder
bands had greater modulation, and the flare is apparent in
some lines (Ne {\sc x} 12 \AA), but not others (O {\sc viii} 19 \AA).

\section{Modeling}
   
We have performed a variable abundance, differential emission measure
(DEM) analysis on line fluxes or upper limits (for details,
see Huenemoerder et al.\ 2001 and references therein).
The DEM is defined as $n_e^2(T) d V(T,n_e)/ dT$, and is a measure of 
the density-weighted plasma volume as a function of temperature and
density. It is, in a sense, the only emergent property derivable from
the integrated line fluxes without imposition of a physical model,
other than the assumptions required to validate the
emissivity model. In our fit to the fluxes, we allowed abundances to
be free parameters, but omitted density sensitive lines and
used a low-density emissivity database. This database (the
Astrophysical Plasma Emission Database, or ``APED'';
Smith et al.\ 2001) tabulates emissivities versus temperature
and density for a plasma in collisional ionization
equilibrium (``CIE''). CIE implies a stable ionization state of a 
thermal plasma under the {\em coronal approximation}, in which the
dominant processes are collisional excitation and ionization from the
ground state, and radiative and dielectronic recombination.  

Model results described below (\S\S\ 3.1, 3.2) were derived from
analysis of the entire observation (i.e., combined quiescent
and flaring time intervals). We also divided the observation
into pre-flare and flare intervals and 
measured the line strengths in each partition. To increase the
potential significance of a temperature diagnostic, we also summed
the Ne~{\sc ix} fluxes and compared to the summed Ne~{\sc x}
flux. In neither the individual lines nor in the summed cases do we
find any significant differences (larger than one standard
deviation) between quiescent and flare states. In
particular, while the elevated high-energy continuum suggests a
modest increase in the 
contribution of high-$T_x$ material to the DEM during the flare, we find that
ratios of the brightest temperature- and density-sensitive
lines did not vary significantly during the course of the
observation (Fig.\ 4). We also find no measurable changes in
abundances between quiescent and flare states.    Hence, for
further analysis, we only considered the entire observation,
in order to derive the physical conditions of the mean state
of TW Hya.

\subsection{Emission Measure Distribution and Abundances}

The emission measure distribution derived from fitting the
Chandra/HETGS spectrum of TW Hya displays a sharp peak at
$\log{T_x} = 6.5$ (Fig.\ 5). The peak at $\log{T_x} \simeq 7.7$ represents
an upper limit and reflects the absence of lines of highly ionized
Fe from the spectrum. The integrated volume emission measure lies within the
range $1.3-1.7\times10^{53}$ cm$^{-3}$. The results are the
same, to within the uncertainties, during both quiescent
and elevated X-ray emission time intervals. The DEM model fitting suggests
the fractional abundances
(relative to solar) of O, Ne, and Fe are $0.3$, $2.0$, and
$0.2$, respectively, while abundances of N, Si, and
Mg are roughly solar. The accuracies of these
derived abundances are about 20\%, as determined from ad hoc
perturbations of the model.

Fig.\ 2 shows a model spectrum predicted from this
``best-fit'' DEM model, convolved with the instrument
response, overlaid on the observed spectrum.
Qualitative agreement of continuum and many lines is good;
density-sensitive features are discrepant, as expected,
since the fit includes only temperature-sensitive lines and
is based on low-density emissivities. 

\subsection{Densities}

The lack of [O {\sc vii}] emission at 22.1 \AA\ in the spectrum of TW Hya (and
the consequent small ratio of the intensity of
this line to that of the 21.8 \AA\ line) implies large
electron densities of $n_e \ge 10^{12}$ cm$^{-3}$ for the
emitting region of this cTTS. The He-like Ne {\sc ix} triplet at
13.45, 13.55, and 13.7 \AA\ provides an even
more stringent density constraint of $\log n_e=12.75$ (cgs; $2\sigma$ limits
12.6--13.0), assuming CIE. We caution that these results
are predicated on the assumption that the EUV continuum from
TW Hya is not driving the 
He-like lines to their high-density ratios (an intense EUV
radiation field can depopulate the metastable level, thereby
weakening the forbidden line and enhancing the
intercombination line; e.g., Porquet \& Dubau
2000). This assumption appears justified, however, as the UV
continuum of TW Hya is relatively weak (though it does
exhibit bright UV line emission; Costa et al. 2000).

The Mg~{\sc xi} He-like triplet is also density
sensitive. However, the
9.0-9.5 \AA\ region shows a curious lack of observed Mg features,
relative to the model.  There are several possible
explanations: the derived abundance could be too high (although this is
unlikely, given the strength of Mg~{\sc xii} 8.42 \AA);
higher-level Lyman transitions of Ne~{\sc x} that are absent from the
APED tables may be present in the Mg~{\sc xi} region; and/or
the model may underestimate the mass of plasma at high
temperatures ($\log{T_x} \geq 7.0$) due to the lack of lines
that are diagnostic of this temperature regime.

\section{Discussion}

\subsection{TW Hya and active, late-type stars: similarities
and differences}

In certain respects, the
Chandra/HETGS spectrum of TW Hya is similar to
that of active, late-type, main sequence or post-main sequence stars, 
for which the abundances of Fe also appear depleted, and the
abundances of Ne appear enhanced, 
with respect to solar (e.g., HR 1099, Drake et al.\ 2001; II Peg,
Huenemoerder et al.\ 2001). Fig.\ 6 compares
Chandra/HETGS spectra of TW Hya and such (primarily)
RS CVn systems in the 13.3 \AA\ to 15.4 \AA\ region, which
encompasses the Ne {\sc IX} 
triplet and several prominent lines of highly ionized
Fe. For the RS CVn binaries the X-ray emission presumably
arises in magnetically confined, coronal plasma. It 
is apparent from this Figure that the Ne to Fe line ratios
of TW Hya are somewhat more extreme than, but not dissimilar
to, those of e.g.\ II Peg and UX Ari. This similarity ---
and, in particular, the prominence of lines of highly ionized Ne ---
could suggest that at least some of the X-ray emission from TW Hya
originates in strong flares during which material is 
evaporated from the stellar photosphere. Evidence for such
time-dependent abundance anomalies has been noted for certain
late-type stars (e.g., Tsuboi et al.\ 1998; Brinkman et al.\
2001; Audard et al.\ 2001; see also discussion in Feigelson 2001).
In the present case, however, there is no evidence that (for
example) the overabundance of Ne is more pronounced while TW Hya is in
a flaring state.

The flare observed during our 
observation of TW Hya displayed very
different behavior from flares seen on  
RS~CVn stars. For example, Huenemoerder et al.\ (2001)
report a flare on II Peg that rose more slowly and
decayed exponentially, consistently with Solar 2-ribbon flares (an arcade of
loops undergoing continuous reconnection). In contrast, the flare
observed on TW Hya displayed a
sharp rise and nearly linear decay. Kastner et al.\ (1999) 
observed flares of similar magnitudes and rise and decay times with ASCA
and noted that the pointed and all-sky survey ROSAT Position Sensitive
Proportional Counter count rates for TW Hya differ by a
factor of two, suggesting that TW Hya routinely flares in this manner.

More importantly, there are significant
differences between the spectra of TW Hya and all
active, late-type stars thus far observed with Chandra/HETGS, 
particularly in the ratios within the density-sensitive 
triplet line complexes of He-like ions (Fig.\ 6).
The large electron densities implied by the O {\sc vii} and
Ne {\sc ix} triplets, $\log{n_e}>12$ (cgs), are 2-3 orders of
magnitude higher than 
those inferred from Chandra/HETGS spectra obtained thus far for active,
late-type (RS~CVn) stars (Canizares et al.\
2000; Huenemoerder et al.\ 2001). It is a long-standing
paradigm that activity in such
stars is due to Solar-like activity scaled up by faster
rotation and the consequent stronger magnetic dynamo (and not,
for instance, close-binary interaction, except via tidally
induced rotation).  TW~Hya has, in constrast, a higher density
and somewhat cooler X-ray emitting region than these ``traditional,''
rotationally-driven dynamo sources.

It must be noted, however, that some stars 
have shown evidence of high coronal
densities. From Fe~{\sc xxi} lines in EUVE
spectra of $\lambda$ And, Sanz-Forcada et
al. (2001) found densities of 12.3 at $\log T=7$, which
increased to 12.9 during a flare.  Brickhouse and Dupree (1998)
reported a very high density, albeit with large
uncertainties, of $\sim$13 for 44i Boo (HD 133640). 

In the Sun, electron densities can be measured for a variety of
coronal structures and temperatures.  In a large flare Phillips et
al.\ (1996a) measured $\log n_e \approx 12$ (cgs) from Fe~{\sc xxi}
lines formed near $\log T = 7$.  Doschek et al.\ (1981) measured flare
densities of $\log n_e$ from 11--12 at the cooler temperatures of
O~{\sc vii}.  In long-duration, gradual (or two-ribbon) flares, Widing
and Doyle (1990) found a logarithmic density of 10.3.  In active
regions, Phillips et al.\ (1996b) determined logarithmic densities of 9-11
from Fe~{\sc xvii} lines, while Kastner \& Bhatia (2001)
determined a logarithmic density of 10 from Fe~{\sc xv} lines.

In sum, a large range of densities is inferred in the Sun and
other stars from high-resolution X-ray and EUV spectra. 
We do not expect a unique value in any case, since we
are sampling emission which is differential in both temperature and
density, and we will determine some mean value of density in the
temperature range over which any particular diagnostic is sensitive.
The important point here is the purely empirical result that at the
temperatures of formation of O~{\sc vii} and Ne~{\sc ix} (6.3 and
6.6 dex, respectively), TW~Hya displays much higher emitting
region densities than those of coronally active stars.  

\subsection{Evidence for and implications of
accretion-powered X-ray emission} 

Given the preceding discussion and the evidence that 
TW Hya likely is surrounded by a
circumstellar disk from which it is still accreting (e.g.,
Muzerolle et al.\ 2000), we consider the possibility that the X-ray
emission originates not in coronal activity but in the
accretion funnel(s) that connects 
the circumstellar disk to the star. The hypothesis that the
X-rays arise in accretion shocks 
along the funnel flows is supported by the emission
measure distribution derived from DEM fitting (Fig.\ 5). The
distribution is sharply peaked, in contrast to the broad
distributions (spanning the range 2 MK to 20 MK)
characteristic of quiescent X-ray emission 
from active, late-type stars (e.g., Huenemoerder et al.\ 2001,
Drake et al.\ 2001). Furthermore the
temperature characterizing the bulk of the plasma,
$\sim3$ MK, is consistent with adiabatic shocks
arising in gas at free-fall velocities of $150-300$ km
s$^{-1}$ --- the range imposed by measurements of the width of
the H$\alpha$ line ($\sim150$ km s$^{-1}$; Muzerolle et al.\
2000) and by the observation that the X-ray emission line
profiles are unresolved by Chandra/HETGS.  We further
note that the volume of X-ray emitting plasma implied by the
derived $n_e$ and emission measure ($\sim1.5\times10^{53}$
cm$^{-3}$) is only of order $10^{-6}$ of the volume of the
star. This fractional volume is far smaller than typical of coronal X-ray
sources (e.g., Huenemoerder et al.\ 2001) and appears more
consistent with that of a ``hot spot'' (or spots) where
accretion impacts the stellar surface. 

If the X-ray emission emanates from such a region, then the densities
inferred from He-like ions can be used to estimate 
the mass accretion rate, given estimates
for the infall velocity (150--300 km s$^{-1}$) 
and filling factor of the accretion
funnel at the surface of the star ($\sim5$\%; Costa et al.\
2000). The inferred densities ($\log{n_e}
\approx 13$) then suggest that the mass accretion rate is
$\sim10^{-8}$ M$_\odot$ yr$^{-1}$. Given the
(considerable) uncertainties, this mass accretion rate is
consistent with the rate suggested by the C IV
$\lambda 1549$\AA\ line flux of TW Hya ($1.2\times10^{-12}$
erg cm$^{-2}$ s$^{-1}$; Valenti et
al.\ 2000), i.e., $\dot{M} \approx 3-6\times10^{-8}
M_\odot$ yr$^{-1}$ (where this estimate is obtained from the
method described in Johns-Krull et al.\ 2000). Both
of these rates are more than an order of magnitude larger than the mass
accretion rate deduced from H$\alpha$ ($\dot{M} \approx
4\times10^{-10} M_\odot$ yr$^{-1}$; Muzerolle et al.\ 2000). 

While significant puzzles remain --- most notably, the
origin of the abundance anomalies observed in its X-ray
spectrum --- these results for TW Hya are consistent with
the hypothesis that most or
all of the X-ray luminosity of this star is derived from
accretion rather than from coronal activity. 
Chandra/HETGS and XMM grating spectroscopy of well-studied,
X-ray-luminous cTTS and wTTS in and near dark clouds, as well as of
the many X-ray-bright wTTS in the TW Hya Association,
would establish whether the physical conditions in
the X-ray emitting region of
TW Hya are representative of TTS and, thereby, might confirm or
refute the notion that accretion contributes to the X-ray
emission from cTTS. High-resolution X-ray spectra of
wTTS will be of particular interest. If analysis of such spectra
yields values of $n_e$ and $T_x$ similar to those found
here for TW Hya, this would call into question the contention that all
wTTS are diskless and/or are no longer accreting material from
their disks.  

\acknowledgements{The authors wish to thank Eric Feigelson,
Andrea Dupree, and Steve Beckwith for incisive comments and
enlightening discussions. This research was supported in part by
contracts SV-61010 and NA-39073 to MIT.}

\newpage

\subsection*{Figure Captions}

\begin{description}

\item[Figure 1] ASCA SIS (CCD) spectrum (top) and Chandra/HETGS
spectrum (bottom). The ASCA counts spectrum has bins of constant
width in energy (29 eV). The Chandra/HETGS counts spectrum has 0.005\AA\ bins 
and has been smoothed by a Gaussian with a width equal to the
instrumental resolution ($\sigma=0.008$\AA). 

\item[Figure 2] Selected regions of the medium energy grating (MEG)
spectrum of TW Hya (solid curve). The observed spectrum is
overlaid with a variable-abundance, differential emission
measure model (dashed curve) that best fits temperature-sensitive
line intensities (\S 3). 

\item[Figure 3] Light curve obtained from the 48 ks Chandra/HETG
observation of TW Hya, integrated over 1.7 \AA\ to 25 \AA\ for
combined HEG and MEG orders -3 to +3 (excluding zero, which was not usable
due to pileup).  Bin sizes are 2 ks.  The background is
negligible. 

\item[Figure 4] Measurements of the ratios of resonance ($r$),
intercombination ($i$), and forbidden ($f$) lines of Ne {\sc ix}
during the quiescent (Q) and flaring (F) states of TW
Hya. Density- and temperature-sensitive diagnostics are
represented by the ratios $f/i$ and $(f+i)/r$,
respectively. Dashed and dotted curves indicate contours of
constant $\log{n_e}$ and $\log{T}$, respectively. The results
indicate that, despite an increase of a factor $\sim2$ in
count rate at the onset of the flare, there was no
measureable change in either $n_e$ or $T$ during the observation.

\item[Figure 5] Differential emission measure (DEM) as a function of 
temperature for TW Hya. This DEM is for the entire
observation and thus represents a mean 
of flare and quiescent states.  It was determined by fitting
low-density limit model emissivities to the observed fluxes for
many elements with their relative abundances as free
parameters.  The hatched region is an upper limit derived from
flux upper limits on undetected high-ionization states of Fe.

\item[Figure 6] Chandra/HETGS spectra of TW Hya and the
active, late-type stars AB Dor (Chandra observation identifier [OID] 16),
Capella (OID 1318), HR 1099 (OID 62538), AR Lac (OID 6, 9),
TY Pyx (OID 601), UX Ari (OID 605), $\lambda$ And (OID 609),
and II Peg (OID 1451) in the spectral region encompassing the 13.56 \AA\
He-like Ne {\sc ix} triplet and the 15.01 \AA\ Fe {\sc XVII} line
(see also Drake et al.\ 2001, Canizares et al.\ 2000, and Huenemoerder
et al.\ 2001). Objects
are RS CVn binaries (or similar systems) with the
exception of TW Hya and the rapidly rotating K dwarf AB Dor. Spectra 
are shown as raw counts in the summed MEG $\pm
1^{st}$ orders, in 0.005 \AA\ bins, and have been
smoothed slightly.  They 
are arranged from top to bottom roughly in order of the Ne {\sc ix}  13.56
to Fe {\sc xvii} 15.01 \AA\ ratio, which is more sensitive to abundance than
to temperature.  Lines labeled are Ne {\sc ix}, Fe {\sc xvii-xviii}, and
O {\sc vii}.

\end{description}

{\small
\begin{deluxetable}{rrrrc}
\tablewidth{0pt}
\tablecaption{TW Hydrae: Selected X-ray Emission Lines}
\tablehead{
\colhead{Line}&
\colhead{$\lambda_t$\tablenotemark{a}}&
\colhead{$\lambda_o$\tablenotemark{b}}&
\colhead{$f_l$\tablenotemark{c}}&
\colhead{$\log T_\mathrm{max}$\tablenotemark{d}}
}
\startdata
Fe \sc    xxv     &   1.8504 &   1.8550 (0.0150)&      (14.9)\tablenotemark{e}& 7.8\\
Si \sc    xiv     &   6.1805 &   6.1745 (0.0087)&       2.7 (1.9) &  7.2   \\
Si \sc    xiii    &   6.6479 &   6.6480 (0.0080)&       2.1 (1.7) &  7.0   \\
Si \sc    xiii    &   6.6882 &   6.6775 (0.0150)&       0.8 (1.2) &  7.0   \\
Si \sc    xiii    &   6.7403 &   6.7400 (0.0052)&       2.8 (1.7) &  7.0   \\
Fe \sc    xxiv    &   7.9857 &   7.9841 (0.0127)&       1.4 (1.4) &  7.3   \\
Mg \sc    xii     &   8.4193 &   8.4157 (0.0062)&       3.0 (2.0) &  7.0   \\
Ne \sc    x       &   9.7081 &   9.7000 (0.0150)&       2.0 (2.1) &  6.8   \\
Ne \sc    x       &   10.239 &  10.2363 (0.0032)&       7.8 (2.8) &  6.8   \\
Fe \sc    xxiv    &   10.619 &  10.6210 (0.0150)&       (5.7)\tablenotemark{e}&  7.3   \\
Fe \sc    xxiv    &   10.664 &  10.6788 (0.0150)&       (5.3)\tablenotemark{e}&  7.3   \\
Ne \sc    ix      &   11.544 &  11.5451 (0.0025)&      18.0 (4.6) &  6.6   \\     
Fe \sc xxii-xxiii &   11.737 &  11.7302 (0.0150)&       (7.4)\tablenotemark{e}&  7.1   \\
Ne \sc    x       &   12.132 &  12.1349 (0.0005)&      73.7 (7.8) &  6.8   \\     
Fe \sc    xvii    &   12.266 &  12.2700 (0.0100)&       4.7 (4.0) &  6.7   \\        
Fe \sc    xx      &   12.820 &  12.8164 (0.0150)&      (13.1)\tablenotemark{e}& 7.0    \\
Fe \sc    xx      &   12.835 &  12.8200 (0.0150)&      (11.7)\tablenotemark{e}& 7.0    \\
Ne \sc    ix      &   13.447 &  13.4508 (0.0008)&     122.4 (11.9) &  6.6   \\     
Ne \sc    ix      &   13.553 &  13.5547 (0.0008)&      79.2 (10.5) &  6.6   \\     
Ne \sc    ix      &   13.699 &  13.7004 (0.0027)&      35.3 (8.6) &  6.6   \\        
Fe \sc    xix     &   13.795 &  13.7898 (0.0150)&       (8.4)\tablenotemark{e}& 6.9    \\
Fe \sc    xvii    &   13.825 &  13.8223 (0.0150)&      (10.0)\tablenotemark{e}& 6.7    \\
Fe \sc    xviii   &   14.210 &  14.2109 (0.0124)&       8.3 (6.9) &  6.8   \\     
Fe \sc    xvii    &   15.014 &  15.0180 (0.0027)&     (179.0)\tablenotemark{e}&  6.7   \\
O  \sc    viii    &   15.176 &  15.1791 (0.0150)&      (15.9)\tablenotemark{e}&  6.7   \\
Fe \sc    xvii    &   15.261 &  15.2605 (0.0026)&      23.2 (8.4) & 6.7    \\     
O  \sc    viii    &   16.006 &  16.0107 (0.0025)&      35.7 (10.4) & 6.5    \\        
Fe \sc    xviii   &   16.073 &  16.0752 (0.0150)&      (17.8)\tablenotemark{e}& 6.7    \\   
Fe \sc    xvii    &   16.780 &  16.7759 (0.0100)&      15.3 (9.9) &  6.7   \\
Fe \sc    xvii    &   17.051 &  17.0511 (0.0048)&      35.9 (13.1) &  6.7   \\
Fe \sc    xvii    &   17.096 &  17.0988 (0.0071)&      25.4 (12.7) &  6.7   \\
O  \sc    vii     &   18.627 &  18.6337 (0.0150)&      10.8 (14.1) &  6.3   \\
O  \sc    viii    &   18.967 &  18.9710 (0.0021)&     195.7 (28.0) &  6.5   \\
O  \sc    vii     &   21.602 &  21.6086 (0.0060)&      91.3 (36.3) &  6.3   \\
O  \sc    vii     &   21.804 &  21.8106 (0.0047)&     105.4 (40.4) &  6.3   \\   
O  \sc    vii     &   22.098 &  22.1225 (0.0150)&      (57.0)\tablenotemark{e}& 6.3  \\
N  \sc    vii     &   24.779 &  24.7859 (0.0050)&      87.8 (36.5) &  6.3   \\
\enddata
\tablenotetext{\ }{Note --- The numbers in parentheses are values of standard deviation.}
\tablenotetext{a}{Theoretical wavelengths of identification, in \AA,
  from APED. If the line is a multiplet, we give the wavelength of the
  stronger component.} 
\tablenotetext{b}{Measured wavelength, in \AA.}
\tablenotetext{c}{Line flux is $10^{-6}$ times the tabulated value in $[\mathrm{phot\,cm^{-2}\,s^{-1}}]$.}
\tablenotetext{d}{Decimal logarithm of temperature [Kelvins] of maximum emissivity.}
\tablenotetext{e}{Line flux is a 2$\sigma$ upper limit.}
\end{deluxetable}




\newpage

\begin{figure*}[htb]
\plotone{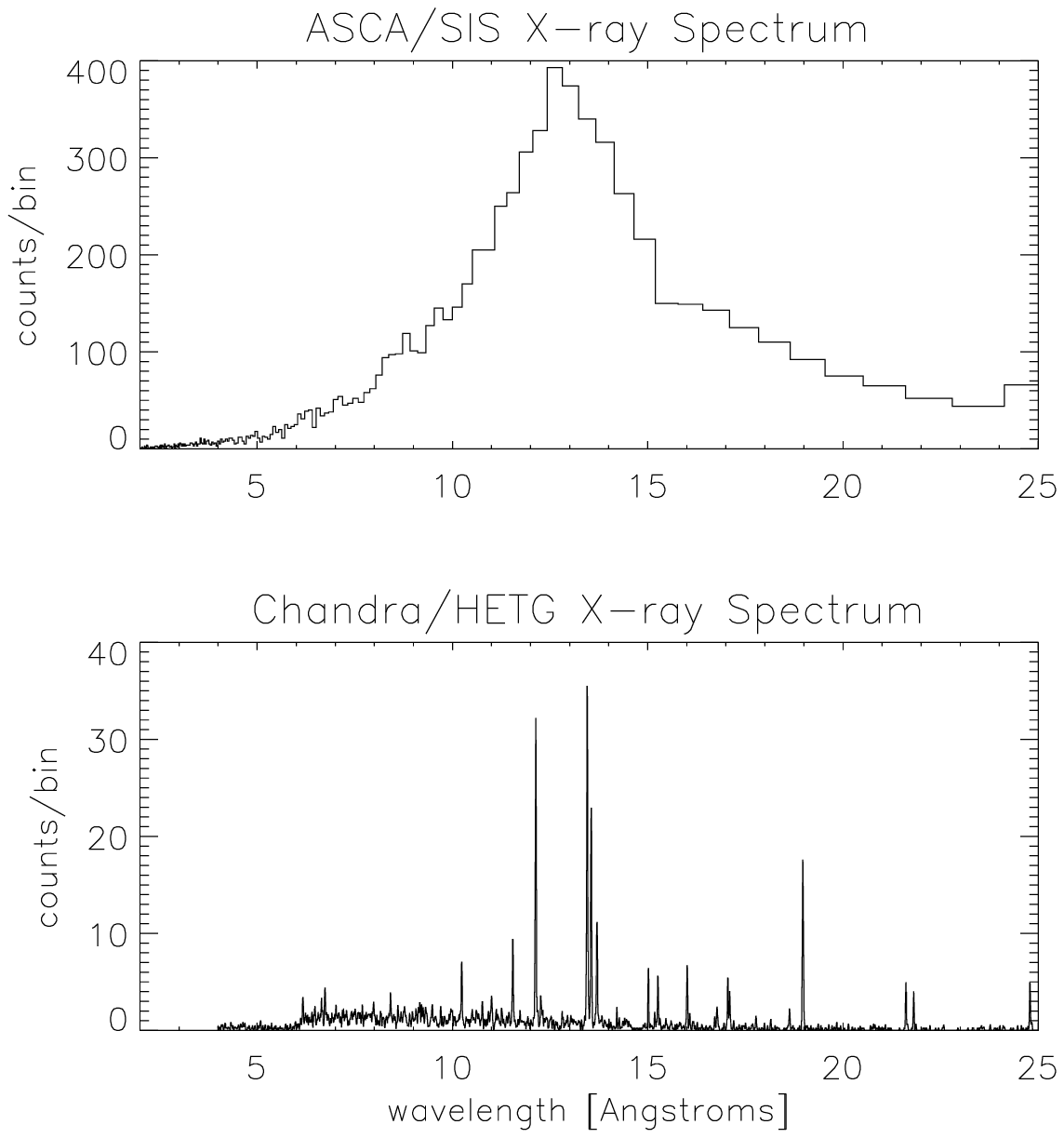}
\end{figure*}

\begin{figure*}[htb]
\includegraphics[scale=0.5,angle=-90]{fig2a.ps}
\includegraphics[scale=0.5,angle=-90]{fig2b.ps}
\includegraphics[scale=0.5,angle=-90]{fig2c.ps}
\end{figure*}

\begin{figure*}[htb]
\plotone{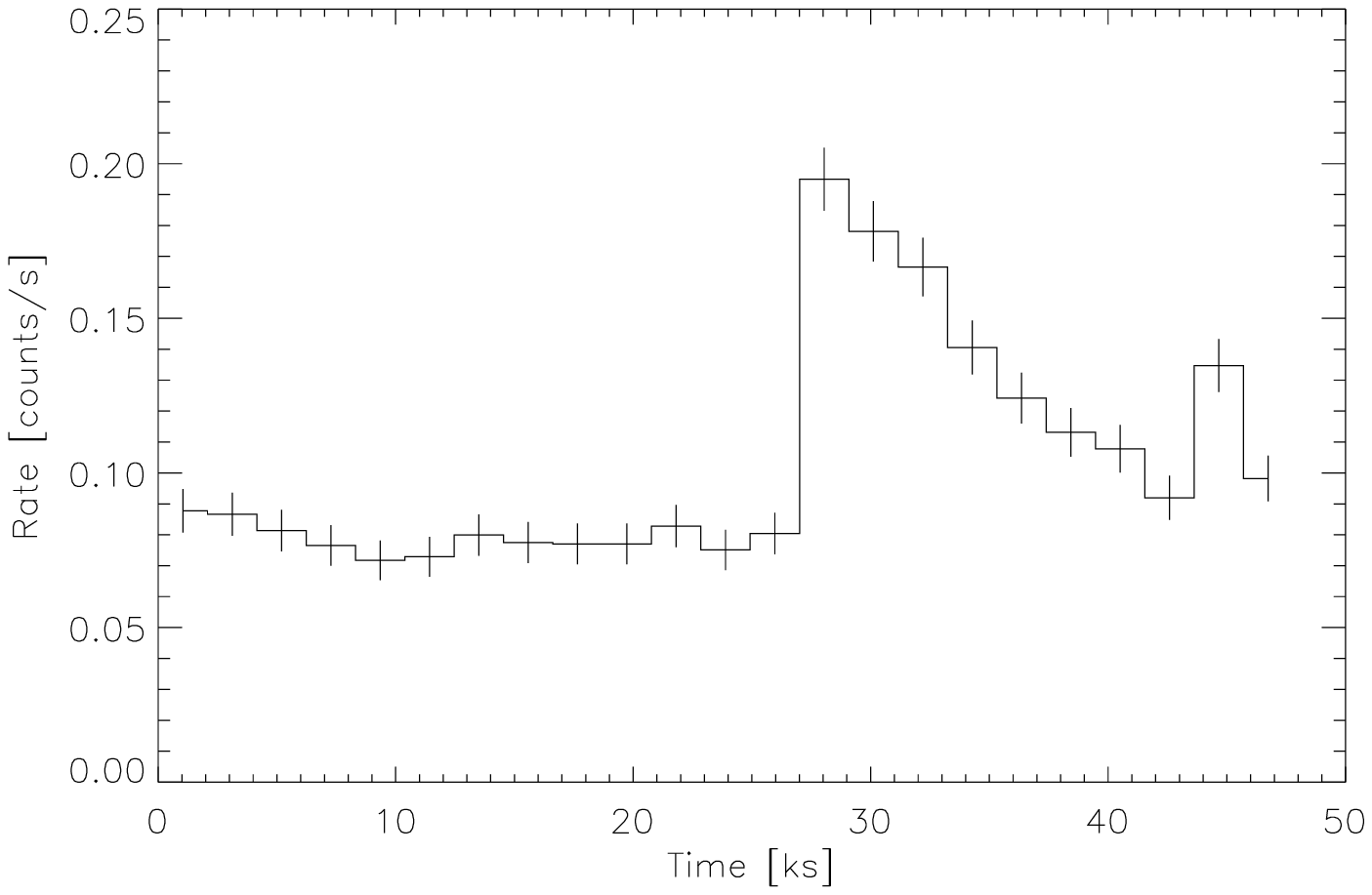}
\end{figure*}

\begin{figure*}[htb]
\plotone{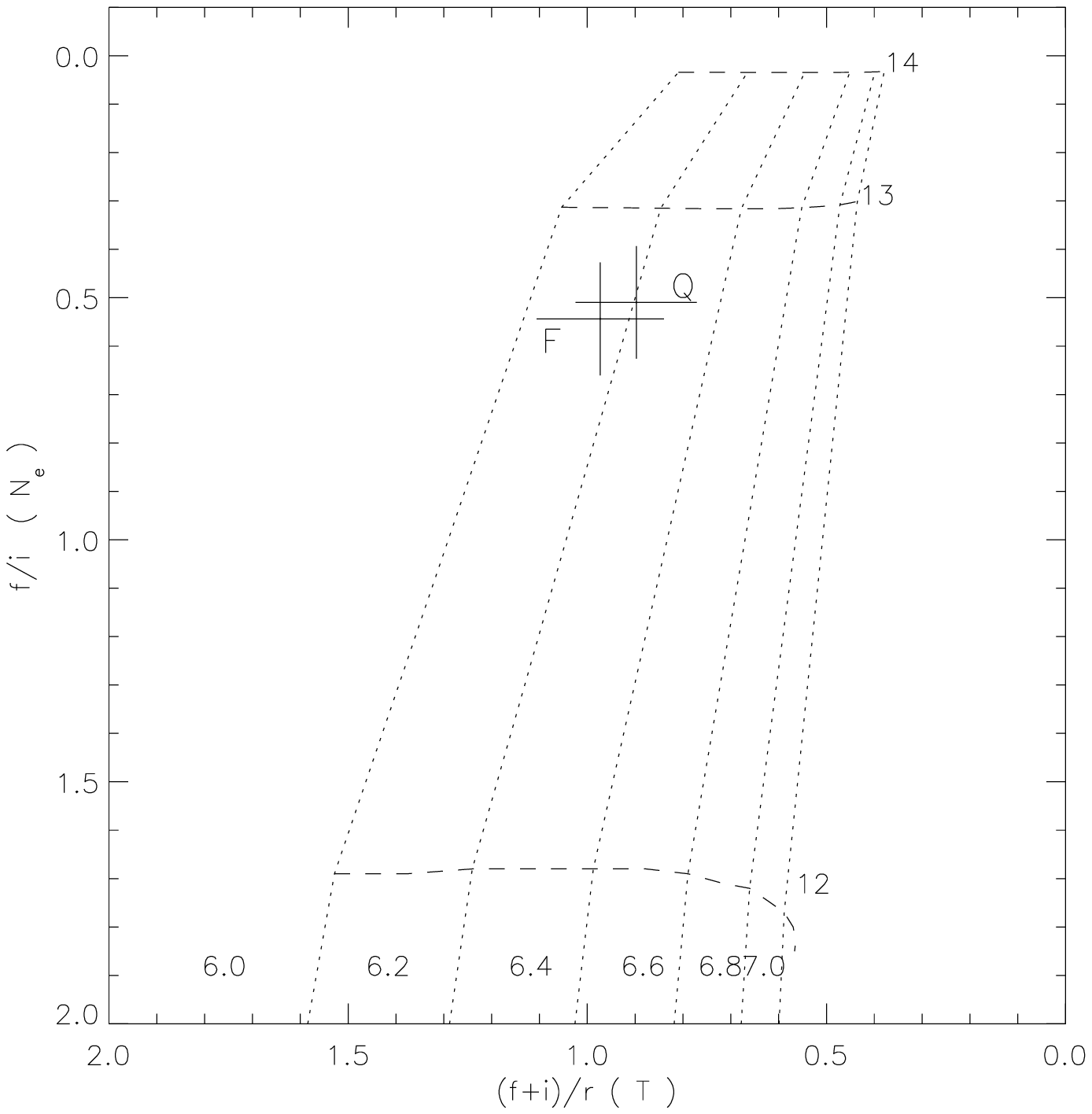}
\end{figure*}

\begin{figure*}[htb]
\plotone{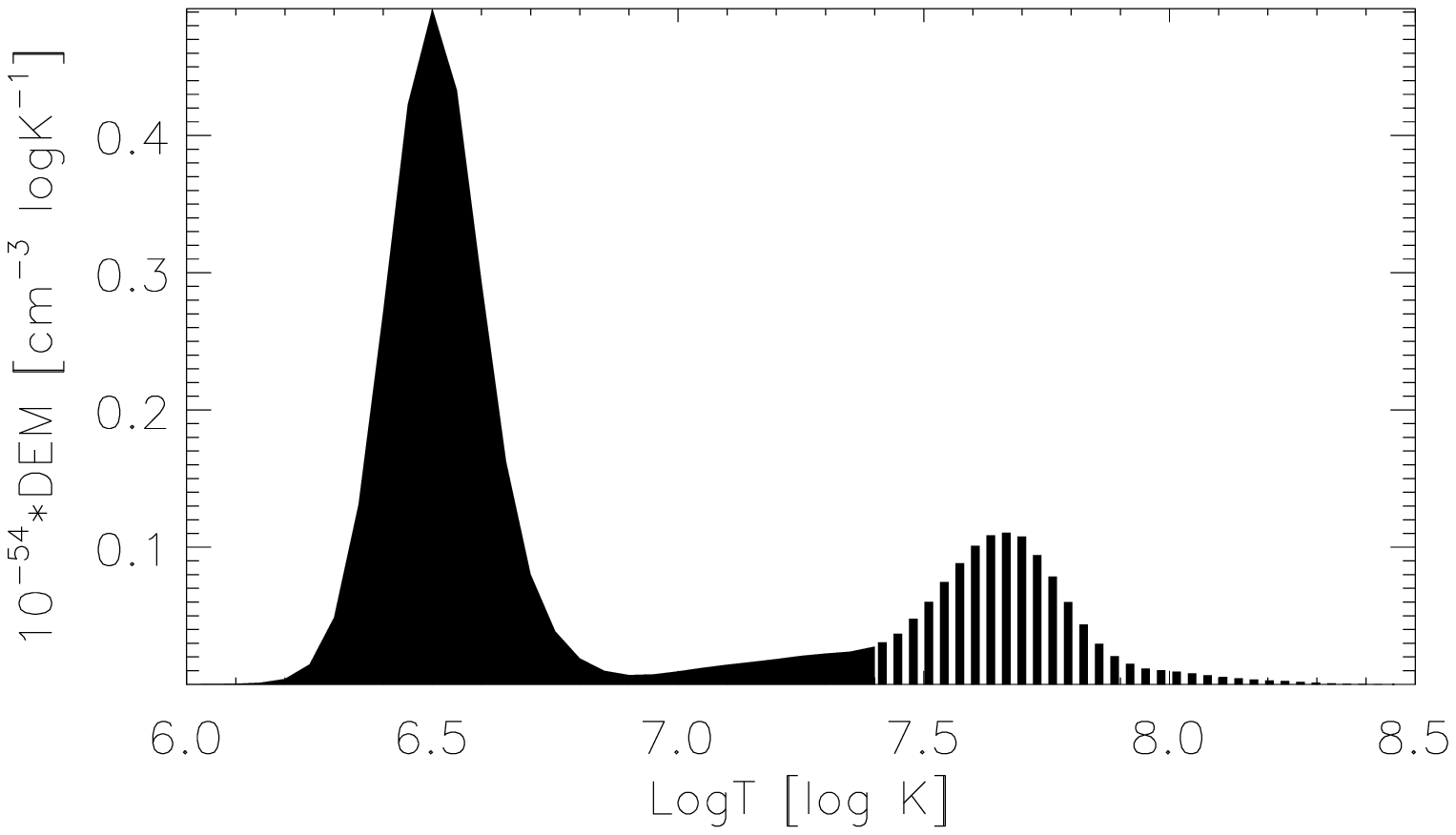}
\end{figure*}

\begin{figure*}[ht]
\plotone{fig6.ps}
\end{figure*}


\end{document}